\documentclass[journal]{IEEEtran}
\usepackage{graphicx}
\usepackage{dcolumn}
\usepackage{bm}
\usepackage{amssymb}
\usepackage{amsfonts}
\usepackage{amsmath}

\newcommand{\Tr}        {\mathrm{Tr}}

\newcommand{\ket}[1]    {\left| #1 \right\rangle}

\newcommand{\kb}[2]     {\left| #1 \right\rangle \!\! \left \langle #2 \right|}

\newcommand{\cM}        {{\mathcal M}}
\newcommand{\cN}        {{\mathcal N}}

\newtheorem{dfn}{Definition}

\begin{document}
\title{Practical Entanglement Distillation Scheme Using Recurrence Method And
 Quantum Low Density Parity Check Codes}
\author{
\IEEEauthorblockN{H.~F. Chau and K.~H. Ho}\\
\IEEEauthorblockA{Department of Physics and Center of Theoretical and
 Computational Physics, \\
 The University of Hong Kong, Pokfulam Road, Hong Kong}}

\maketitle

\begin{abstract}
 Many entanglement distillation schemes use either universal random hashing or
 breeding as their final step to obtain almost perfect shared EPR pairs.
 In spite of a high yield, the hardness of
 decoding a random linear code makes the use of random hashing and breeding
 infeasible in practice.  In this pilot study, we analyze the performance of
 the recurrence
 method, a well-known entanglement distillation scheme, with its final random
 hashing or breeding procedure being replaced by various efficiently decodable
 quantum codes.  Among all the replacements investigated, the one
 using a certain adaptive quantum low density parity check (QLDPC) code is
 found to give the highest yield for Werner states over a wide range of noise
 level --- the yield for using this QLDPC code is higher than the first runner
 up by more than 25\% over a wide parameter range.  In this respect, the
 effectiveness of using QLDPC codes in practical entanglement distillation is
 illustrated.
\end{abstract}


\IEEEpeerreviewmaketitle
\section{Introduction}\label{Sec:Intro}

 Bipartite entanglement distillation, or entanglement distillation for short,
 describes a general class of methods for obtaining copies of high fidelity
 Bell states shared by two
 parties from a collection of initially shared entangled particles using local
 quantum operations and classical communication.  Many entanglement
 distillation methods have been developed.  For instance, two cooperative
 players, Alice and Bob, may compare their measured error syndromes of their
 shares of the quantum particles using a pre-determined quantum
 error-correcting code (QECC) and then perform the necessary error recoveries.
 Alternatively, they may apply an entanglement distillation (ED)
 procedure such as the well-known recurrence
 method~\cite{BBP+96a,BDSW96a}.
 More precisely, by two-way local operations and classical communications
 (LOCC2), Alice and Bob discard those particles whose measurement results are
 not consistent with that of the corresponding Bell states.  Thus, two-way ED
 can be regarded as a carefully designed quantum-error-detection-code-based
 error rejection method.

 QECC- and ED-based entanglement distillation methods can be extended in many
 ways such as the introduction of
 adaptation~\cite{Chau02,GL03a,MS00,LS07a,LS07b} and various breeding
 methods~\cite{VV05,HDM06,HDM06b}.  Most studies so far focus on improving
 the yield or the maximum error-tolerable rate under various conditions.
 Most of them use either the random hashing method introduced in
 Ref.~\cite{BDSW96a} or variations of the breeding method proposed in
 Refs.~\cite{BBP+96a} and~\cite{VV05} as their last step.  (There are a few
 exceptions such as the distillation methods reported by Chau~\cite{Chau02} and
 Cirac \emph{et al.}~\cite{CEM99}.  However, the yield of the former drops
 dramatically as the noise level of the initially shared EPR pairs increases
 and the latter gives at most one EPR pair at the end.)

 The use of random hashing or breeding as the last step of entanglement
 distillation is not surprising.  Both
 methods make use of random stabilizer quantum error-correcting codes whose
 error syndromes can be measured using a simple and efficient quantum circuit
 plus a few ancillas.  Also, Vollbrecht and Verstraete found that the optimal
 yield protocol to distill identical copies of high fidelity Werner states
 among those using local unitary operations and Bell basis measurements
 uses a certain carefully designed breeding method~\cite{VV05}.
 Moreover, Fig.~3 in Ref.~\cite{VV05} shows that their optimal protocol
 narrowly beats the universal random hashing method when the fidelity of the
 Werner
 state is high.  Third, the yields of both random hashing and breeding methods
 are easily computable.  Finally, in the limit of an arbitrarily large number
 of shared qubits remaining before applying the random hashing or the breeding
 procedure, the fidelity of the finally distilled EPR pairs, if any, will be
 arbitrarily close to one.  These nice features will not be present
 if the universal
 random hashing or breeding procedures are replaced by a general easily
 decodable code of finite length.  Nonetheless, this is what Alice and Bob have
 to do in order to make the distillation method practical because decoding
 random linear codes (associated with the hashing or breeding protocol) is an
 NP-complete problem~\cite{BMT78,FCG+88}.

 Among the entanglement distillation methods in the literature, one of the
 easiest and most well-known is the recurrence method, which can tolerate up to
 50\% quantum error rate upon repeated application~\cite{BBP+96a,BDSW96a}.  In
 this pilot study, we focus on the yield of applying the recurrence method with
 the final random hashing or breeding procedures being replaced by various
 efficiently decodable QECCs (of finite length).  We also
 investigate a more aggressive strategy in which the efficiently decodable QECC
 is adaptively chosen according to error syndrome measurement
 results as they become available.  The efficiently decodable QECC
 is used adaptively either to
 correct erroneous qubits like a QECC-based scheme or to reject erroneous ones
 like an ED-based scheme.  Since the fidelity of the distilled EPR pairs may no
 longer be arbitrarily close to one, to analyze the yield, we have to compute
 the number of perfect EPR pairs given that the quantum error rate of the
 distilled pairs is less than or equal to a certain threshold.  Actually, this
 criterion, which is a straightforward generalization of a similar criterion
 used in the study of classical codes~\cite{BS01}, has already been adopted by
 MacKay \emph{et al.} in their study of the performance of certain quantum low
 density parity check (QLDPC) codes~\cite{MMM04a}.  We find that among all the
 codes we
 have investigated and over almost all ranges of initial error rates for the
 Werner states, the yield of the resulting ED procedure for a certain novel
 adaptive
 QLDPC stabilizer code is better than the yields of all the other efficiently
 decodable codes we have investigated (by at least 25\%).

 We begin by stating the detailed procedure of our recurrence-method-based ED
 procedure and the figure of merit used in our study in Sec.~\ref{Sec:Aim}.
 Since our ED procedure may make use of a novel QLDPC code, we spend the
 whole of Sec.~\ref{Sec:Scheme} discussing the rationale behind using this
 code as well as its construction and efficient decoding.  We also write down a
 detailed procedure of how to use the QLDPC code as one of the possible final
 steps to distill entanglement in an adaptive way there.  Then we study the
 performance of our scheme to distill almost perfect EPR pairs from a set of
 Werner states in Sec.~\ref{Sec:Num}.  In particular, among all the practical
 methods we have studied, the best yield under most circumstances is obtained
 by replacing the computationally intractable random hashing or breeding
 methods by a certain adaptive QLDPC code.  Finally, we conclude in
 Sec.~\ref{Sec:Conc} by suggesting some possible future works.

\section{Entanglement Distillation Method Used And Figure Of Merit}
\label{Sec:Aim}

 We study the performance of repeated rounds of the recurrence
 method~\cite{BBP+96a,BDSW96a} combined with an efficiently decodable code.
 More precisely, Alice and Bob apply $r$ rounds of the recurrence method for
 some $r =
 0,1,2,\ldots$ before finally applying an efficiently decodable code.  And in
 each round of recurrence, they randomly pair up their remaining share of
 qubits and measure the syndrome of the parity check $X\otimes X$, $Y\otimes Y$
 or $Z\otimes Z$ for each pair where the unitary operations $X,Y,Z$ are given
 by
\begin{subequations}
\begin{equation}
 X = \left[ \begin{array}{ll} 0 & 1 \\ 1 & 0 \end{array} \right] ,
 \label{X_def}
\end{equation}
\begin{equation}
 Y = \left[ \begin{array}{ll} 0 & -i \\ i & 0 \end{array} \right] ,
 \label{Y_def}
\end{equation}
 and
\begin{equation}
 Z = \left[ \begin{array}{ll} 1 & 0 \\ 0 & -1 \end{array} \right] ,
 \label{Z_def}
\end{equation}
\end{subequations}
 respectively.  They keep the remaining particles in each pair only if their
 measured syndrome is consistent with that given by two perfect EPR pairs each
 in the state $|\Psi^+\rangle \equiv (|00\rangle+|11\rangle)/\sqrt{2}$.
 Whereas for the efficiently decodable codes, we choose the followings in this
 study:
 the $[[(4^t-1)/3,(4^t-1)/3-2t,3]]$ quantum Hamming codes over the field
 $GF(4)$, the $[2^t-1,2^t-1-t,3]$ classical Hamming codes over $GF(2)$, the
 $[[2^t-1,2^t-1-2t,3]]$ Calderbank-Shor-Steane (CSS) codes constructed from the
 classical Hamming codes over the field $GF(2)$, the $[2^t,2^t-2t,5]$
 two-error-correcting BCH classical code, the $[2t+1,1,2t+1]$ classical
 majority vote code, and a $(8,16)$-regular QLDPC code with codeword size $960$
 to be reported in Sec.~\ref{Sec:Scheme}.  (Note that we include a few
 classical codes in this study because occasionally it is more effective to use
 them for part of the quantum error correction procedure.  For instance, the
 last step used in the most error tolerant two-way prepare-and-measure-based
 quantum key distribution scheme known to date uses the classical majority vote
 code~\cite{Chau02}.  A few high performance degenerate quantum codes
 are constructed by concatenating classical codes with quantum
 codes~\cite{DSS98,SS07}.) Our collection of efficiently decodable codes
 studied includes a few high rate quantum and classical codes that correct one
 to two quantum or classical errors.  We do not include
 multiple-error-correcting quantum codes because either their rate is low or
 their decoding method is complicated.

 As we have mentioned in Sec.~\ref{Sec:Intro}, the average fidelity of the
 resulting shared $|\Psi^+\rangle$'s distilled by the above methods cannot be
 arbitrarily close to one.  And we follow MacKay \emph{et al.} by demanding
 that the quantum error rate of the distilled $|\Psi^+\rangle$'s be less
 than or equal to a fixed small threshold parameter
 $p_\textrm{th}$~\cite{MMM04a}.
 Fortunately, we find that the conclusions of our study do not sensitively
 depend on the choice of $p_\textrm{th}$; and for simplicity, we set
 $p_\textrm{th} = 2.0\times 10^{-5}$.  In fact, this choice is consistent with
 the bit error rate commonly used to compare the performance of classical
 error-correcting codes~\cite{BS01}.  Note that our choice of the threshold
 quantum error rate $p_\textrm{th}$ implies that the entropy of the distilled
 $|\Phi^+\rangle$'s must be less than or equal to $-(1-p_\textrm{th}) \log_2
 (1-p_\textrm{th}) - p_\textrm{th} \log_2 (p_\textrm{th} / 3) \approx 3.7\times
 10^{-4}$.

 The yield of a scheme is defined as the number of shared perfect
 $|\Psi^+\rangle$'s distilled divided by the number of initially shared
 imperfect pairs in the limit of an arbitrarily large number of initially
 shared pairs.  And for simplicity, we study the yield in the event that each
 initially shared pair is in the Werner state
\begin{eqnarray}
 W_{p_0} & = & (1 - p_0) \kb{\Psi^+}{\Psi^+} + \frac{p_0}{3} \left(
  \kb{\Psi^-}{\Psi^-} \right. + \nonumber \\
 & & \left.  \kb{\Phi^+}{\Phi^+} + \kb{\Phi^-}{\Phi^-} \right) ,
\end{eqnarray}
 where $p_0$ is the quantum error rate of the initially shared pairs and
 $|\Psi^\pm\rangle \equiv (|00\rangle\pm|11\rangle)/\sqrt{2}$,
 $|\Phi^\pm\rangle \equiv (|01\rangle\pm|10\rangle)/\sqrt{2}$ are the four
 maximally entangled Bell states.

 For a fixed value of $p_0$, we may maximize the yield of our
 recurrence-method-based ED procedure by varying the number of rounds of
 recurrence and the actual parity check used in each round.  And we use this
 optimized yield $D$ (or yield for short in case there is no ambiguity
 possible) as the figure of merit to compare the efficiency of different
 efficiently decodable codes in the last step of our
 recurrence-method-based ED procedure.  Finally, we emphasize that the number
 of rounds of recurrence and the actual parity check used in each round is a
 function of $p_0$ and the final efficiently decodable code used.

\section{Adaptive Quantum Low Density Parity Check Code And Belief
 Propagation Decoding}
\label{Sec:Scheme}

\subsection{Why Use Adaptive QLDPC Code}\label{Subsec:Rationale}

 The rationale behind using adaptive QLDPC codes in an ED procedure is simple.
 Since both QECC- and ED-based schemes use QECC codes, it is instructive to
 investigate methods that can adaptively locate the erroneous qubits and
 perform the necessary error recovery or error rejection.  That is to say,
 Alice and Bob first pick a stabilizer QECC ${\mathcal C}$ and they decide to
 do one of the followings according to their measured syndromes:
 1)~to discard or to apply error correction to a particular qubit;
 or 2)~to replace
 ${\mathcal C}$ by a compatible stabilizer code ${\mathcal C}'$ formed by
 appending a few more parity checks to ${\mathcal C}$ so as to hope to more
 precisely locate the erroneous qubits at the expense of a lower yield.

 QLDPC codes are particularly suited for this purpose for a number of
 reasons.  First, their parity check matrices are sparse so that their average
 error-correcting capabilities do not in general change greatly with the
 deletion of a few qubits.  Second, QLDPC codes can be efficiently
 constructed~\cite{MMM04a,COT05a} and efficient approximate decoding algorithms
 such as belief propagation for classical low density parity check codes (or
 LDPC codes for short)~\cite{P88,MM98,Mac99} can be readily extended to QLDPC
 codes~\cite{HC08,PC08}.  Finally, families of compatible QLDPC codes exist
 and are easily constructible~\cite{MMM04a,HC08}.  For this reason, we use
 them to replace the random hashing code in entanglement
 distillation.

\subsection{Constructing Quantum Low Density Parity Check Codes}
\label{Subsec:QLDPC}

 A QLDPC code can be defined and represented in a way very similar to a
 classical LDPC code.

\par\medskip
\begin{dfn}
 A quantum low density parity check (QLDPC) code is a quantum stabilizer block
 error-correcting code over a finite field $GF(q)$ that has a sparse parity
 check matrix.  In particular, a $(d_v, d_c)$-regular QLDPC code has a sparse
 parity check matrix $H$ with a constant column weight $d_v$ and a constant row
 weight $d_c$~\cite{MMM04a, COT05a}.
\end{dfn}

\par\medskip
 For example, by explicitly writing down their parity check matrices, one can
 see that the quantum error-detection-code associated with each round of
 recurrence method~\cite{BBP+96a,BDSW96a} and the Leung and Shor
 method~\cite{LS07a, LS07b} are $(1,2)$- and $(2,4)$-regular QLDPC codes,
 respectively.  (In some sense, these two codes are atypical QLDPC codes as
 they are composed of tensor products of block codes of sizes $2$ and $4$,
 respectively.)  Actually, a large number of QLDPC codes exist for a
 sufficiently large block code size.  Existing ways to construct them include
 the so-called bicycle and unicycle constructions by MacKay
 \emph{et al.}~\cite{MMM04a}, group theoretical construction by Camara
 \emph{et al.}~\cite{COT05a}, algebraic-combinatorial-based construction of
 quasi-cyclic CSS QLDPC code by Hagiwara and Imai~\cite{HI07}, classical
 quasi-cyclic LDPC-based construction of QLDPC codes by Hsieh
 \emph{et al.}~\cite{HBD08a}, finite geometric construction by
 Aly~\cite{SAA07a}, and BCH- and finite geometry LDPC code-based asymmetric
 QLDPC codes construction by Sarvepalli \emph{et al.}~\cite{SRK08a}.  It is
 remarkable that the error-correcting capability of some of these QLDPC codes
 are better than the Gilbert-Varshamov rate for binary CSS codes~\cite{MMM04a}.

 The $GF(4)$ stabilizer QLDPC code we used in this study is a simple but
 important extension of the bicycle construction by MacKay \emph{et al.} in
 Ref.~\cite{MMM04a}.  Actually, our QLDPC code construction works for any
 $q$-ary code where $q = p^v$ is a prime power.  So, we report this more
 general construction below.  This construction is based on our earlier
 unpublished work~\cite{HC08}.

 We follow the notation of Ashikhmin and Knill~\cite{AK01} by defining the
 unitary operators $X_a$ and $Z_b$ acting on a $q$-dimensional Hilbert space by 
\begin{equation}
 X_a: \ket{i} \longmapsto \ket{i+a} \label{X_a_def}
\end{equation}
and
\begin{equation}
 Z_b: \ket{i} \longmapsto \varpi_p^{\Tr(ib)}\ket{i} \label{Z_b_def}
\end{equation}
 for all $a, b, i \in GF(q)$, where $\varpi_p$ is the $p$th root of unity and 
\begin{equation}
 \Tr (i) = i + i^p + \cdots + i^{p^{v-1}} \in GF(p) \label{trace_def}
\end{equation}
 is the absolute trace of $i\in GF(q)$.  Note that all arithmetic inside the
 state ket and in the exponent of $\varpi_p$ is performed in the finite
 field $GF(q)$.  We also identify the unitary operator $X_a Z_b$ with 
 $a + b \omega_{q^2} \in GF(q^2)$ where $\omega_{q^2}$ is a fixed primitive
 element in $GF(q^2)$.  Using this identification, we may abuse our language by 
 saying, for example, that a qudit has experienced an error $a + b
 \omega_{q^2}$.

 To construct a $[n,n-k]$ bicycle $GF(q^2)$ stabilizer QLDPC code with row
 weight $d_c$, we first select a random $(n/2) \times (n/2)$ cyclic $GF(q^2)$
 sparse-matrix $C_{\textrm{B}}$ with row weight $d_c/2$.  That is to say, the
 elements of the matrix $C_{\textrm{B}}$ satisfy $(C_{\textrm{B}})_{i,j} 
 \equiv (C_{\textrm{B}})_{ij} = \alpha_{i -j}$ for some $\alpha_{i-j} \in
 GF(q^2)$, where $\left( \alpha_i \right)_{i=1}^{n/2}$ is a sparse vector with
 weight $d_c/2$.  So, for $1\leq i, i', j \leq n/2$,
\begin{equation}
 (C_\textrm{B})^T_{i, i + i' -j} = (C_\textrm{B})_{i',j} \label{C_transpose1}
\end{equation}
and 
\begin{equation}
 (C_\textrm{B})^T_{i', i + i' -j} = (C_\textrm{B})_{i,j} , \label{C_transpose2}
\end{equation}
 where $C_{\textrm{B}}^T$ denotes the transpose of $C_{\textrm{B}}$.  We
 define the $(n/2) \times n$ bicyclic matrix $H$ by
\begin{equation}\label{eq:H_bi}
 H = [C_{\textrm{B}}, C_{\textrm{B}}^T] .
\end{equation}
 Clearly, rows of this bicyclic matrix $H$ are mutually orthogonal to each
 other with respected to the skew-symmetric inner product
\begin{equation}\label{eq:in_prod}
 (a + b \omega_{q^2} | c + d \omega_{q^2}) \equiv \Tr ( ad - bc) \in GF(p)
\end{equation}
 for all $a, b, c ,d \in GF(q)$, irrespective of whether $C_B$ is sparse
 or not.  Since 
\begin{equation}
 X_c Z_d X_a Z_b = \varpi_p^{(a+ b \omega_{q^2}| c + d \omega_{q^2})}
 X_a Z_b X_c Z_d , \label{phase_X_Z}
\end{equation}
 the rows of $H$ can be identified as the generators of the stabilizer of a
 $q$-ary QECC~\cite{AK01,CRSS98a}; and so is $H_\textrm{B}$, the matrix
 obtained by deleting a few rows of $H$.  In this way, $H_\textrm{B}$ becomes
 the parity check matrix of a $q$-ary QLDPC code.  More importantly, the
 $GF(q^2)$ QLDPC code constructed in this way is not necessarily a CSS code. 

 Interestingly, we may build a large number of regular QLDPC codes using this
 modified bicycle construction.  The trick is to pick the sparse vector 
 $(\alpha_i)_{i=1}^{n/2}$ in such a way that
\begin{equation} \label{eq:con1}
 |\{ i : \alpha_{n'i + j} \neq 0 \}| = u
\end{equation}
 for all $j$ with the constraint that $(n / 2)$ is divisible by $n'$, where the
 symbol $|\cdot|$ denotes the number of elements in the argument set.  That is
 to say, we pick the sparse vector $(\alpha_{i'})_{i'=1}^{n/2}$ in such a way
 that all the length $n / (2 n')$ sub-vectors obtained by picking the $i$th
 components with $i = j \pmod{n'}$ have the same Hamming weight $u$.  Then it
 is easy to check that the parity check matrix $H$ constructed is $(n'u,
 2n'u)$-regular.  And, by deleting the $(in' + j)$th row of $H$ for $i \in
 \mathbb{N}$, $j \in J$ where $J$ is a proper subset of $\{1, 2, \cdots, n'\}$,
 the resulting parity check matrix $H_B$ corresponds to a $([n' - |J|]u, 2 n'
 u)$-regular $q$-ary QLDPC code.  Note that the proportion of rows in $H$ that
 are removed to obtain $H_B$ equals $|J|/n'$.  Thus, $H_B$ has $n$ columns and
 $n (n' - |J|)/ (2n')$ rows so that it encodes $n - n(n' - |J|)/ (2n')$ qudits.
 For instance, let $q = 2, n = 12, n' = 3, (\alpha_i) = (1, \omega_4,
 \omega_4^2, 0, 0, 0)$ where $\omega_4$ is a primitive element in $GF(4)$, and
 $J = \{ 3 \}$.  Then our construction gives the $(2, 6)$-regular binary QLDPC
 stabilizer (but non-CSS) code
\begin{equation}
 \left[
 \begin{array}{cccccccccccc}
 1 &\omega_4 &\omega_4^2 &0 &0 &0 & 1 & 0 & 0 & 0 &\omega_4^2 & \omega_4 \\
 0 &1 &\omega_4 &\omega_4^2 &0 &0 & \omega_4 &1 & 0 & 0 & 0 &\omega_4^2 \\
 0& 0 & 0 &1 &\omega_4 &\omega_4^2 &0 & \omega_4^2 &\omega_4 & 1 & 0 & 0 \\
 \omega_4^2& 0 & 0 & 0 &1 &\omega_4 &0 & 0  & \omega_4^2 &\omega_4 & 1 & 0 
 \end{array}
 \right]. \label{QLDPC_example}
\end{equation}

 Recall that for a $(d_v, d_c)$-regular QLDPC code with codeword length $n$,
 the number of non-zero elements in its parity check matrix $H$ is $d_v n$.
 This number is also equal to the number of rows of $H$ times $d_c$.  Thus, the
 (quantum) rate of the $(d_v, d_c)$-regular QLDPC code is greater than or equal
 to $1 - d_v / d_c$, where the equality holds if any only if the rows of $H$
 are linearly independent over $GF(q)$.  In our subsequent study, we only
 consider those $H$'s with full row rank so that their rate is equal to $ 1 -
 d_v/ d_c$.  Surely, this extra constraint on the choice of $H$ is not very
 restrictive as our construction is likely to give $H$ with full row rank
 anyway.

 Note that for a typical sparse vector $(\alpha_i)_{i = 1}^{n/2}$ satisfying
 Eq.~(\ref{eq:con1}), the number $|\{i: \alpha_i = \beta \}| / (n/2)$ is about
 the same for all $\beta \in GF(q^2) \setminus \{ 0 \}$.  To summarize, we have
 succeeded in constructing a large number of regular $q$-ary QLDPC codes. The
 construction is simple and efficient:  We need a (pseudo-)random
 number generator to generate the sparse vector $(\alpha_i)_{i=1}^{n/2}$.
 And the stabilizer of our modified bicycle construction can be specified
 from $(\alpha_i)_{i=1}^{n/2}$ and the information regarding which rows to
 delete.  Thus, at most $\mbox{O}(n)$ classical bits of storage space are
 required to specify our regular QLDPC codes.  Note further that the almost
 equal probability of occurrence of non-zero elements in
 $(\alpha_i)_{i=1}^{n/2}$ and the regularity of our QLDPC codes are two of the
 reasons why our QLDPC codes are reasonably effective to combat quantum errors.

\subsection{Belief Propagation Algorithm For Quantum Stabilizer Codes}
\label{Subsec:BP}

 Similar to classical LDPC code, a QLDPC code can be represented by the
 so-called Tanner graph~\cite{HC08,PC08}.  Let $\mathcal{C}$ be a QLDPC code
 with parity check matrix $H$ encoding each $k$ qudits of information as an $n$
 qudit state.  Its associated Tanner graph is a bipartite graph with vertex set
 $V = V_1 \cup V_2$.  Each variable node in $V_1$ is associated with a qudit of
 the code represented by a column of $H$; and each check node in $V_2$ is
 associated with a generator of the code represented by a row of $H$. 
 There is an edge linking $i \in V_2$ and $j \in V_1$ if and only if $H_{ij}
 \neq 0$.

 Many efficient approximate decoding strategies for classical LDPC codes can be
 regarded as message passing algorithms executed on the corresponding Tanner
 graph.  Famous for its linear runtime in the codeword size $n$ provided that
 the error probability of each bit is independent, belief propagation is one of
 the most commonly used message passing algorithm in which the messages passed
 between nodes in a Tanner graph are conditional
 probabilities~\cite{P88,MM98,Mac99}.  More importantly, belief propagation
 algorithm is also applicable to quantum stabilizer codes whose generators of
 the stabilizer is sparse.  Moreover, its efficiency also scales linearly with
 $n$ in case the error probability of each qudit is independent.  Actually, a
 similar decoding scheme for QLDPC and for graph states can be found in
 Refs.~\cite{PC08} and~\cite{LP08}, respectively.  And the presentation below
 is adapted from our earlier unpublished manuscript~\cite{HC08}.

 Since the belief propagation algorithm is also applicable to $GF(q^2)$
 stabilizer codes, we explicitly write down this more general situation here.
 By passing the messages between the nodes, the task of the belief propagation
 decoding algorithm is to infer a tentative decoding
 $\tilde{\textrm{\boldmath$x$}}$.  That is to say,
 $\tilde{\textrm{\boldmath$x$}}$ is the most likely value of error experienced
 by the shared EPR pairs based on the measured error syndrome vector 
\begin{equation}
 \textrm{\boldmath$s$} \equiv (s_i)_{i \in V_2} =
 (\sum_{j \in V_1} (H_{ij} | e_j))_{i \in V_2}, \label{syndrome_vector}
\end{equation}
 where the check node $s_i \in GF(p)$ is the $i$th component of the syndrome
 $\textrm{\boldmath$s$}$, $e_j$ is the error experienced by the variable node
 $x_j$, and $(\cdot | \cdot)$ is the skew-symmetric inner product defined in
 Eq.~(\ref{eq:in_prod}).  We call $\textrm{\boldmath $e$} \equiv (e_j)_{j \in
 V_1}$ the noise vector of the state shared by the sender and the receiver. 

 The messages consist of two types of conditional probabilities $Q_{ij}^\alpha$
 and $R_{ij}^\alpha$ associated with each non-zero entry in the parity check
 matrix $H$ for all $\alpha \in GF(q^2)$.  To aid discussions, we call
 the $j$th component of the tentative decoding vector 
 $\tilde{\textrm{\boldmath$x$}}$ the variable node
 $\tilde{\textrm{\boldmath$x$}}_j \in GF(q^2)$.
 The quantity $Q_{ij}^\alpha$ approximates the belief that the qudit 
 $\tilde{\textrm{\boldmath$x$}}_j$ has experienced the error  
 $\alpha \in GF(q^2)$ given the messages received from all its checks other
 than $i$.  And the quantity $R_{ij}^\alpha$ is the probability
 of check $i$ being satisfied given that the variable node 
 $\tilde{\textrm{\boldmath$x$}}_j$ has experienced an error in the state
 $\alpha \in GF(q^2)$ and the components of $\tilde{\textrm{\boldmath$x$}}$
 other than $\tilde{\textrm{\boldmath$x$}}_j$ have a separable distribution
 given by the probabilities $Q_{ij}^\alpha$'s. 

 Initially, each message $Q_{ij}^\alpha$ is set to the prior probability
 $f_j^\alpha$ that $x_j$ has experienced an error $\alpha$. In situation of our
 interest, $f_j^\alpha$ is a quantity of the quantum channel linking the two
 parties who would like to perform entanglement distillation.  In each step, 
 the quantities $R_{ij}^\alpha$ are updated according to the equation
\begin{equation}\label{eq:Rij}
 R_{ij}^\alpha = \sum_{\textrm{\boldmath $x$}':x'_j=\alpha} \left[
 {\rm Pr}(s_i|\textrm{\boldmath $x$}')
 \prod_{j' \in \cN(i)  \setminus\{j\} } Q_{ij'}^{x'_{j'}} \right] ,
\end{equation}
 where $\cN(i) \equiv \{ j:H_{ij} \neq 0 \}$ denotes the set of variable nodes
 participating in the check $i$ and
\begin{equation}
 \textrm{Pr}(s_i|\textrm{\boldmath $x$}') = \left \{
 \begin{array}{ll}
 1 & \mbox{if $\textrm{\boldmath $x$}'$ satisfies the check $i$}, \\
 0 & \mbox{otherwise}.
 \end{array} \right. \label{conditional_Pr_1}
\end{equation}
 That is to say,
\begin{eqnarray}
 \textrm{Pr} (s_i | \textrm{\boldmath $x'$}) 
 &=& \delta \left( \sum_{j' \in V_1} (
 H_{ij'} | x'_{j'}), s_i \right)\nonumber \\
 &=& \delta \left( \sum_{j' \in \cN(i) \setminus \{j\}} (
 H_{ij'} | x'_{j'}), s_i - (H_{ij} | \alpha ) \right) \label{Conditional_Pr_2}
\end{eqnarray}
 where 
\begin{equation}
 \delta(x, y) = \left \{
 \begin{array}{ll}
 1 & \mbox{if $x = y$}, \\
 0 & \mbox{otherwise},  
 \end{array}
 \right. \label{delta_def}
\end{equation}
 is the Kronecker delta.

 For QLDPC stabilizer codes, Eq.~(\ref{eq:Rij}) 
 can be computed efficiently using a fast-Fourier-transform-like recursive
 iteration.  In other words, we observe that 
\begin{equation}\label{eq:R}
 R_{ij}^\alpha = R_{ij; \cN(i) \setminus \{j\}, s_i -(s_{ij}|\alpha)}
\end{equation}
 where
\begin{equation}
 R_{ij;J,b} = \sum_{ \{ x'_{j'} :  j' \in J \}}
 \left[\delta\left( \sum_{j' \in J} \left(H_{ij'}|x'_{j'} \right),b \right)
 \prod_{j' \in J} Q_{ij'}^{x_{j'}'}\right] \label{Sum_Product_1}
\end{equation}
 for all $b\in GF(p)$.  Then we can evaluate Eq.~(\ref{eq:Rij}) by recursively
 applying the identity
\begin{equation}
 R_{ij;J,b} = \sum_{c \in GF(p)} R_{ij;J_1,c}
 R_{ij;J_2,b - c} \label{Sum_Product_2}
\end{equation}
 for any partition $\{J_1, J_2 \}$ of the set $J$ with $|J_1| \approx |J_2|$
 until $|J| = 1$. (And surely for $J = \{ j' \}$, $R_{ij;J,b}$ can be
 calculated directly using Eq.~(\ref{eq:R}).)

 After computing $R_{ij}^\alpha$ efficiently,
 each check node $s_i$ sends the message $R_{ij}^\alpha$ to the variable node
 $x_j$ for all $j \in \cN(i)$.  Next, each variable node updates the messages
\begin{equation}\label{eq:Q1}
 Q_{ij}^\alpha = \phi_{ij} f_j^\alpha \prod_{i' \in \mathcal{M}(j)\setminus
 \{i\}} R_{i' j}^\alpha
\end{equation}
 according to the information $R_{i' j}^\alpha$'s from check nodes $s_{i'}$'s
 for all $i' \in \cM(j) \setminus \{ i\}$,
 where $\cM(j) \equiv \{ i:H_{ij} \neq 0 \}$ is the set of checks involving
 variable node $x_j$. The normalization constants $\phi_{ij}$'s ensure that
 the sum of conditional probabilities $\sum_{\alpha \in GF(q^2)} Q_{ij}^\alpha
 = 1$. 

 After each round of message passing, we compute the pseudo-posterior
 probabilities
\begin{equation}\label{eq:Q2}
 Q_j^\alpha =  \phi_j f^\alpha_j\prod_{i \in \mathcal{M} (j)} R^\alpha_{ij},
\end{equation}
 where $\phi_j$ is a normalization constant making $\sum_\alpha Q_j^\alpha =
 1$.  We now set $\tilde{x}_j$, the $j$th component of 
 the tentative decoding $\tilde{\textrm{\boldmath $x$}}$, 
 to $\alpha$ if $Q_j^\alpha \geq Q_j^\beta$ for all $b\in GF(q^2)$.  And we
 denote this operation by
\begin{equation}\label{eq:argmax}
 \tilde{x}_j = \raisebox{-1ex}
 {\mbox{$\stackrel{\textrm{argmax}}{\scriptstyle \alpha \in GF(q^2)}$}}~
 Q_j^\alpha.
\end{equation}

 The decoding algorithm iterates until either the tentative decoding
 $\tilde{\textrm{\boldmath $x$}}$ is consistent with the observed syndrome
 (that is, $s_i = \sum_{j \in V_1}(H_{ij}|\tilde{x}_j ) $ for all $i \in V_2$) 
 or a pre-determined maximum rounds of message passing is reached.

 To summarize, the belief propagation algorithm can be
 applied to decode QECC codes because its decisions depend only on our prior
 assumptions of the noise of the channel and the measurement results for an
 independent noise channel of the
 error syndrome.  Moreover, it decodes QLDPC codes efficiently partly because
 each summand in Eq.~(\ref{eq:Rij}) can be expressed as a sum of products.

\subsection{Detailed Procedure Of Using Adaptive Quantum Low Density Parity
 Check Code}

 The detailed procedure of using adaptive QLDPC code to distill the final
 EPR pair is shown below.  Our procedure is based on a much more general
 framework of using adaptive QLDPC code to distill entanglement (reported
 in our unpublished work in Ref.~\cite{HC08}).

\par\medskip
\begin{enumerate}
[{[Adaptive Entanglement Distillation Using Quantum Low Density Parity Check
 Codes And Belief Propagation]}]
\item
 Alice and Bob randomly pick an $(8,16)$-regular QLDPC code $H[2]$ with
 codeword size $960$ using our generalization of MacKay \emph{et al.}'s bicycle
 construction reported in Sec.~\ref{Subsec:QLDPC}.  By deleting a few parity
 checks from $H[2]$ using the method reported in Sec.~\ref{Subsec:QLDPC}, they
 obtain a $(6,16)$-regular QLDPC code $H[1]$, which is a subcode of $H[2]$.
\item \label{al:measure}
 Alice and Bob measure their corresponding shares of the noisy EPR pairs using 
 the QLDPC code $H[1]$ with the help of (unentangled) ancillas.  Alice sends
 her measurement results to Bob.  And then Bob computes the error syndrome
 $\textrm{\boldmath$s$}[1]({\textrm{\boldmath $e$}})$, where
 ${\textrm{\boldmath $e$}}$ is the noise vector of the state they shared.
\item \label{al:bp}
 Using the belief propagation algorithm and Eq.~(\ref{eq:Q2}), Bob computes the
 posterior marginal probabilities $Q_j^\alpha [1]$ that his $j$th qubit has
 experienced an error $\alpha \in GF(4)$ given the messages passed from all its
 check nodes.  From the posterior marginal probabilities, Bob deduces a
 tentative decoding $\tilde{\textrm{\boldmath$x$}} [1]$ based on the
 measured error syndrome
 $\textrm{\boldmath$s$}[1]({\textrm{\boldmath $e$}})$.

\begin{figure}[t]
 \centering
 \includegraphics*[scale = 0.34]{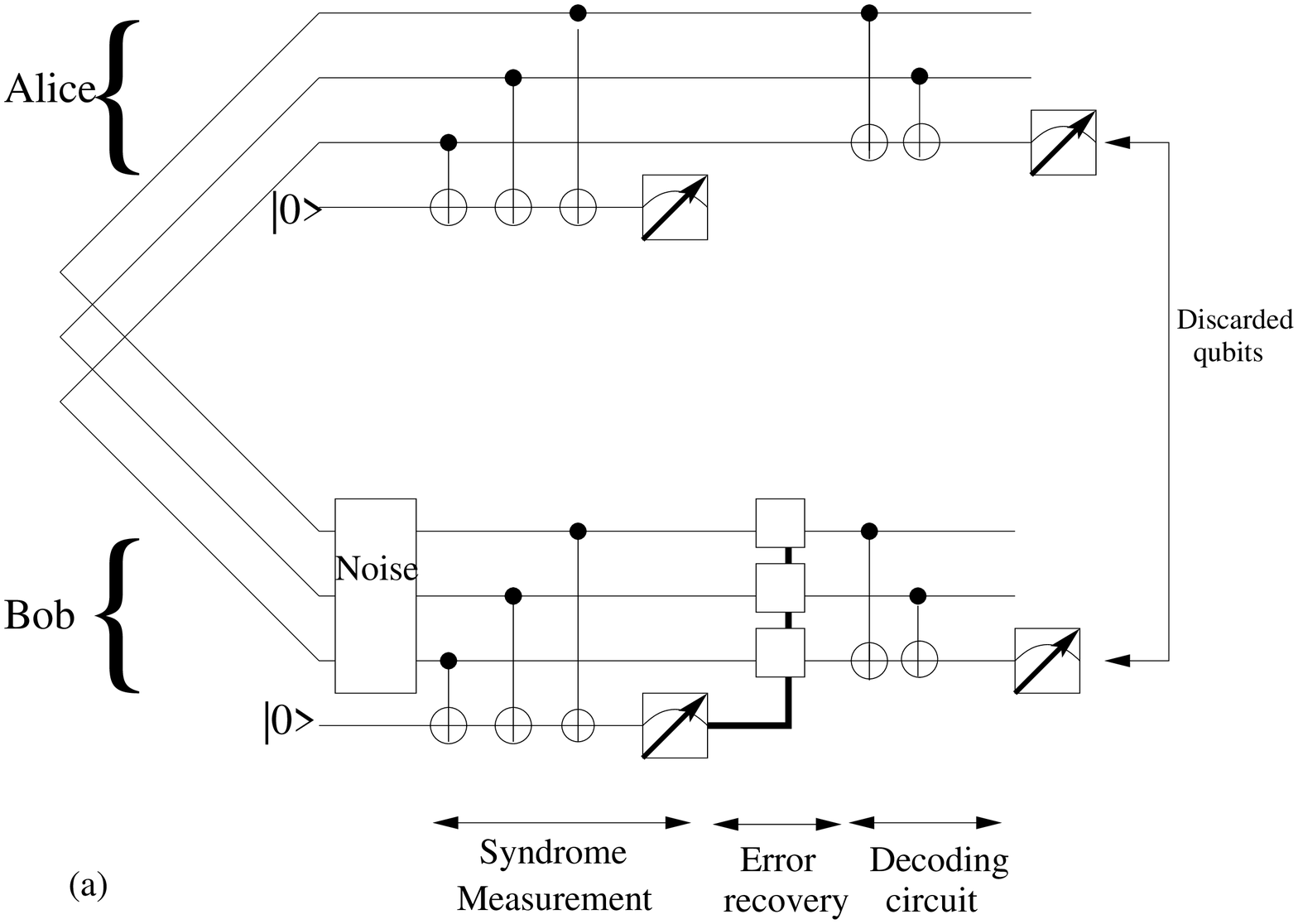}
 \includegraphics*[scale = 0.34]{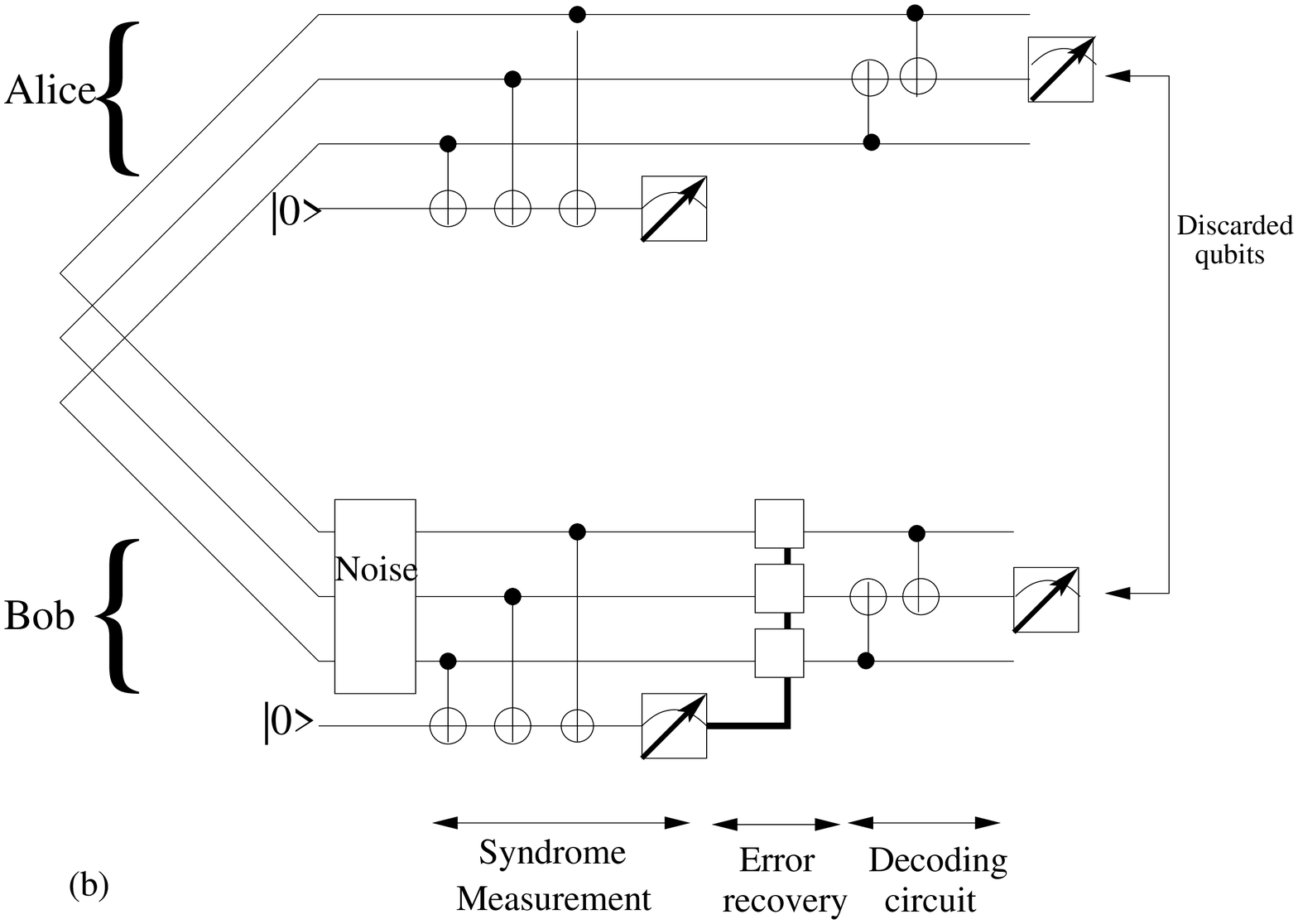}
 \caption{An illustrative example of adaptive entanglement distillation using
  the code $H[1] = (\omega_4 \; \omega_4 \; \omega_4)$.  In case the error can
  be corrected by $H[1]$, decoding circuits in (a) and (b) are equivalent up to
  permutation of entangled qubits.  But in case the error cannot be corrected
  by $H[1]$, the results of the two decoding circuits may differ due to the
  error propagation in the decoding process.}
\label{fig:de}
\end{figure}

\item \label{al:bp2}
 If a tentative decoding $\tilde{\textrm{\boldmath$x$}} [1]$ satisfying
 $H[1] \tilde{\textrm{\boldmath$x$}}[1] = \textrm{\boldmath$s$}[1]
 ({\textrm{\boldmath $e$}})$ is found within the first
 $m_{\textrm{max}} = 5$ rounds of message passing, then
 $\tilde{\textrm{\boldmath$x$}}[1]$ is also a self-consistent error vector.
 (Just like the case of decoding classical QLDPC codes using belief
 propagation algorithm, the choice of $m_{\textrm{max}}$ does not sensitively
 affect the performance provided that it is of order of unity.)
 In this case, what Bob needs to do is to perform the error
 correction by applying the additive inverse of the pseudo-posterior noise
 vector, namely $-\tilde{\textrm{\boldmath$x$}} [1]$, to his qubits.
 The resulting state is likely to be copies of almost perfect encoded EPR
 pairs.  Finally, Alice and Bob finish up by running the encoding circuit for
 $H[1]$ backward to distill copies of almost perfect EPR pair.
 (See Fig.~\ref{fig:de}a.)  This marks the end of our scheme.
\item \label{th:bp}
 If $H[1] \tilde{\textrm{\boldmath$x$}}[1] \neq
 \textrm{\boldmath$s$}[1] ({\textrm{\boldmath $e$}})$ even after
 $m_{\textrm{max}}$ rounds of belief propagation message passing, then Alice
 and Bob substitute the QLDPC code $H[2]$ for $H[1]$ and repeat
 steps~\ref{al:measure}--\ref{al:bp2} again.
 If a tentative decoding still cannot be found (that is,
 $H[2] \tilde{\textrm{\boldmath$x$}}[2] \neq
 \textrm{\boldmath$s$}[2] ({\textrm{\boldmath $e$}})$ even after
 $m_{\textrm{max}}$ rounds of belief propagation message passing), then Alice
 and Bob discard those EPR pairs whose beliefs of finding valid decodings are
 low.  More precisely, they throw away the $j$th EPR pair if the entropy of the
 pseudo-posterior probabilities
 \begin{eqnarray}
  h_4(Q_j [2])
  & \equiv & h_4(\{ Q_j^\alpha [2] : \alpha \in GF(4) \} ) \nonumber \\
  & = & -\sum_{\alpha \in GF(4)} Q_j^\alpha [2] \log_2 Q_j^\alpha [2]
  \label{E:entropy_Q_j}
 \end{eqnarray}
 is greater than the entropy threshold $h_\textrm{th} = S(W_{p_0}) = -(1-p_0)
 \log (1-p_0) - p_0 \log (p_0/3)$.

 The detailed procedure to throw away a EPR pair requires attention.  According
 to the belief propagation algorithm, Alice and
 Bob believe that the most probable error experienced by the $j$th EPR pair is 
 $\alpha_j[2] = \{\alpha[2] \in GF(4): Q_j^\alpha[2]
 \geq Q_j^\beta[2], \forall \beta \in GF(4) \}$.  So Bob first applies
 $-\alpha_j[2]$ to his share of the $j$th EPR pair.  Surely, there is more
 than one possible encoding circuit for $H[2]$ and running any of these
 encoding circuits backward can correctly decode $H[2]$ in the absence of
 noise.  Since the tentative decoding cannot be found, in order to minimize the
 decoding error, Alice and Bob run the encoding circuit backward in which the
 sum of the entropies of the pseudo-posterior probabilities for the message
 qubits are minimized.  After applying this decoding circuit, they can throw
 away those shared EPR pairs with high entropy of the pseudo-posterior
 probabilities. (See Fig.~\ref{fig:de}.)
\end{enumerate}

\par\medskip
 The choice of codes and parameters in the above distillation procedure
 requires explanation.  Note that the rate of $H[1]$, which equals
 $5/8$, is quite high.  And at the same time, our Monte Carlo simulation shows
 that it is very effective to correct most errors when the quantum error rate
 is less than about 1\%.  Thus, $H[1]$ is our main ingredient of obtaining a
 high yield.  In the event that tentative decoding
 $\tilde{\textrm{\boldmath $x$}}[1]$ cannot be found, by switching to the
 code $H[2]$, which has a higher error-correcting capability at the expense of
 a relatively low rate of $1/2$, we hope to maintain a reasonable yield without
 lowering the fidelity of the resulting EPR pairs too much.  And in the worst
 possible situation that the tentative decoding
 $\tilde{\textrm{\boldmath $x$}}[2]$ cannot be found, we switch to a much lower
 yield ED-like method.  We hope that this adaptive combination of QLDPC codes
 combined with the recurrence method can effectively and efficiently distill
 EPR pairs.

 Surely, the above method is very flexible and can be easily generalized and
 modified.  For instance, one may consider using multiple levels of QLDPC codes
 each with very different rates.  Unfortunately, the average error-correcting
 capability of a QLDPC code with large codeword size can only be numerically
 simulated to date.  That is why various authors use numerical simulations to
 evaluate the performance of QLDPC
 codes~\cite{MMM04a,COT05a,HI07,HBD08a,SRK08a}.  Thus, the search for the
 highest yield recurrence method with the final random hashing being replaced
 by an efficiently decodable adaptive quantum code is a very difficult task.
 Nonetheless, we have tried to play around with a few efficiently decodable
 codes and find that the currently reported one gives the highest yield. 

\section{Yield Of Practical Entanglement Distillation Using Recurrence Method}
\label{Sec:Num}

 We study the (optimal) yield of our recurrence-method-based ED procedure
 using a few efficiently decodable codes to distill a collection of Werner
 states $W_{p_0}$.  Here the term optimal refers to the maximum yield
 obtained by tuning the number of rounds of recurrence and the measurement
 basis used in each round of recurrence.
 For all but the adaptive QLDPC code we have studied, the
 optimal yield $D$ can be calculated analytically.  While for the adaptive
 QLDPC code, the yield is computed based on Monte Carlo simulations.  Actually,
 we focus on the final step by studying the performance of the adaptive QLDPC
 code by Monte Carlo simulations.  And once this performance is established,
 the optimal yield $D$ can be deduced in the same way as other efficiently
 decodable block codes.  We use the following method to find out the
 performance of the adaptive QLDPC code.  For a given quantum error rate of the
 input noise vector, we numerically compute the resultant quantum error rate
 and the (partial) yield when we feed these noisy qubits into the adaptive
 QLDPC code with belief propagation decoding and then run the encoding circuit
 backward.  To obtain a reliable estimate, we take the average over $5\times
 10^7$ independently generated noise vectors for each input quantum error rate.
 After performing these Monte Carlo simulations, we find that up to an error of
 2\% or less, the resultant quantum error rate of the qubits $e_\textrm{out}$
 and the partial yield $D_\textrm{partial}$ are given by:
\begin{equation}
 \log_{10} \left( e_\textrm{out} \right) \approx -5.01 + 93.70 e_\textrm{in}
 \label{eq:approx_performance1}
\end{equation}
 and
\begin{equation}
 D_\textrm{partial} \approx 0.628 - 0.0032 e^{205.3 e_\textrm{in}} ,
 \label{eq:approx_performance2}
\end{equation}
 whenever the input quantum error rate $e_\textrm{in}$ is in the interval
 $[0.0025,0.015]$.

\begin{figure}[t]
 \centering
 \includegraphics*[scale = 0.44]{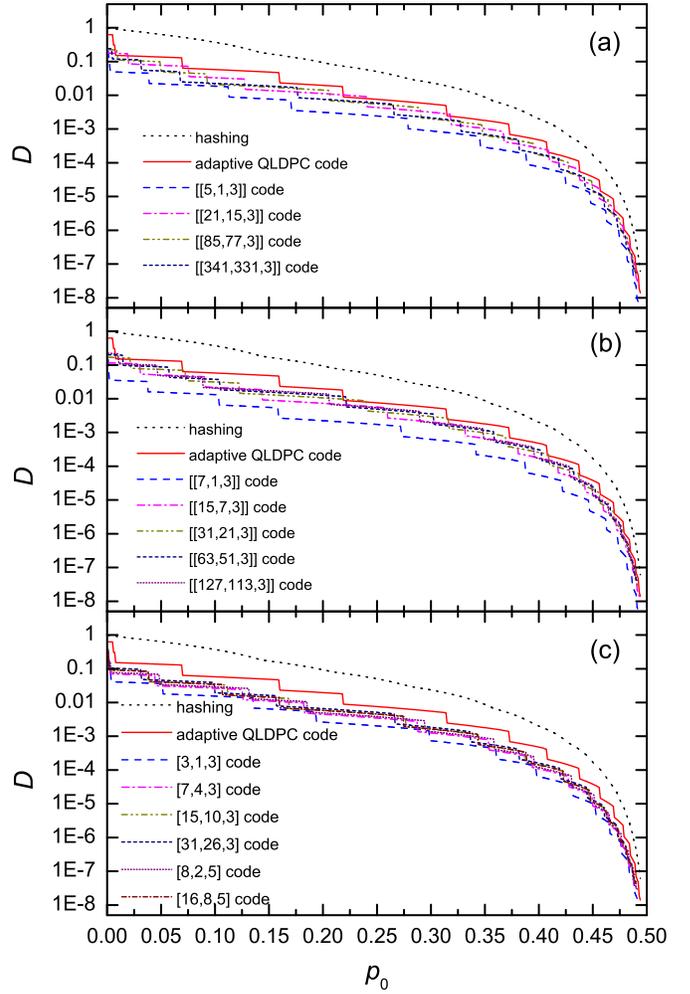}
 \caption{The yield $D$ of the recurrence method using various efficiently
  decodable codes as the last step to distill a collection of Werner states
  $W_{p_0}$ as a function of $p_0$.  The quantum error rate of the distilled
  EPR pairs is less than or equal to $p_\textrm{th} = 2.0\times 10^{-5}$.  The
  solid curve is the case of using the adaptive $(8,16)$-regular QLDPC code
  with codeword size $960$ and the dotted curve is the case of using random
  hashing plotted for comparison.  Fig.~(a) and~(b) show also the yields for a
  few quantum $GF(4)$ Hamming codes and CSS codes constructed from classical
  $GF(2)$ Hamming codes, respectively.  Fig.~(c) depicts also the yields for
  the $[3,1,3]$ classical majority vote code, a few classical $GF(2)$ Hamming
  codes and the $[16,8,5]$ BCH code.}
\label{fig:yield-log}
\end{figure}

 Fig.~\ref{fig:yield-log} shows the yields $D$ as a function of $p_0$ when the
 standard random hashing code is replaced by a number of different efficiently
 decodable codes and when the quantum error rate of the distilled EPR pairs is
 less than $p_\textrm{th} = 2.0\times 10^{-5}$.  In other words, the entropy
 of the distilled EPR pairs is at most $3.7\times 10^{-4}$.  (Note that the
 yield is about the same when the standard random hashing procedure is replaced
 by the so-called two-copy breeding proposed introduced by Vollbrecht and
 Verstraete in Ref.~\cite{VV05}.  So, we do not show it in
 Fig.~\ref{fig:yield-log} to avoid overcrowding the graph.  Since the
 error of using Eqs.~(\ref{eq:approx_performance1})
 and~(\ref{eq:approx_performance2}) to evaluate the performance of the adaptive
 $(8,16)$-regular QLDPC is about 2\%, we decide not show the corresponding
 error bars in the figure.)

 One distinctive feature of the $p_0-D$ curves are that, except for the case of
 using random hashing (and also breeding), they are discontinuous in various
 places.  Since all the efficiently decodable codes we have investigated are
 not universal, each discontinuity in the $p_0-D$ curve corresponds to a change
 in the number of recurrence used in order to achieve the optimal yield.  In
 contrast, the $p_0-D$ curve for the universal random hashing is continuous
 with several cusps.  Each cusp corresponds to a change in the number of
 recurrence used in order to achieve the optimal yield.

 Another observation is that the yield when using random hashing (and also
 breeding) outperforms all the other codes we have studied.  This is not
 surprising because the rates of all the other codes we have investigated are
 lower than that of random hashing and breeding.  In fact, this is the price
 Alice and Bob have to pay in order to make the entanglement distillation
 scheme practical.

\begin{figure}[t]
 \centering
 \includegraphics*[scale = 0.44]{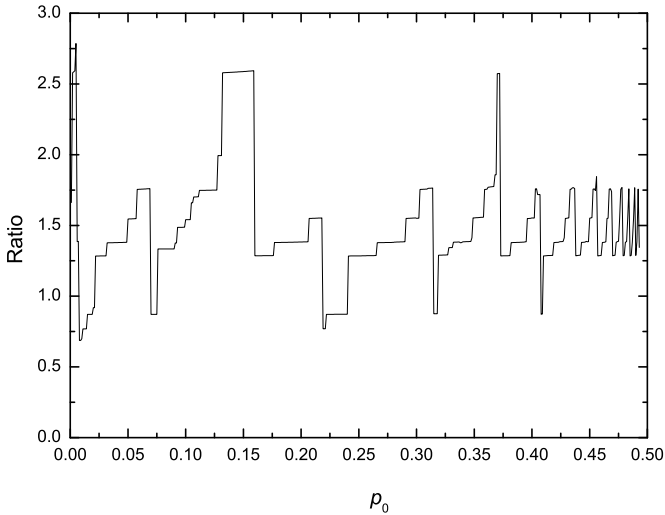}
 \caption{The ratio of the yield for using the $(8,16)$-regular QLDPC code
  with codeword size $960$ to the best yield using other efficiently decodable
  codes selected in this study against $p_0$.}
\label{fig:yield-ratio}
\end{figure}

 The most important observation deduced from Fig.~\ref{fig:yield-log} is that
 the yield using the adaptive $(8,16)$-regular QLDPC code outperforms all the
 other efficiently decodable codes we have investigated.  To show the
 performance of the adaptive $(8,16)$-regular QLDPC code in a clearer way, we
 use the same data in Fig.~\ref{fig:yield-log} to plot a new graph.  This new
 graph (Fig.~\ref{fig:yield-ratio}) plots the ratio of the yield using the
 adaptive $(8,16)$-regular QLDPC code to the best yield using other efficiently
 decodable codes we have investigated against $p_0$.  Figs.~\ref{fig:yield-log}
 and~\ref{fig:yield-ratio} show that over a wide range of values of $p_0$, the
 yield of our adaptive $(8,16)$-regular QLDPC code with $n = 960$ is at least
 25\% better although occasionally the $[[21,15,3]]$, $[[85,77,3]]$ $GF(4)$
 quantum Hamming codes and the $[[31,21,3]]$, $[[63,51,3]]$ and $[[127,113,3]]$
 CSS codes constructed from the corresponding $GF(2)$ classical Hamming codes
 win.  Moreover, Fig.~\ref{fig:yield-log} shows that no classical code we have
 studied outperforms the adaptive $(8,16)$-regular QLDPC code in terms of the
 yield $D$.  These discoveries suggest that efficiently decodable degenerate
 codes may not give a high yield when used with recurrence method.
 Fig.~\ref{fig:yield-log} also depicts that although the yields of using any
 family of codes, such as the $[[(4^t-1)/3,(4^t-1)/3-2t,3]]$ $GF(4)$ quantum
 Hamming codes generally decreases as $t$ increases, there are plenty of
 exceptions violating this general trend.  This complexity originates from the
 complicated interplay between two opposing effects, namely, as $t$ increases,
 the code rate increases while the quantum error rate of particles after
 applying the code reduces less effectively.  In this respect, it seems rather
 unlikely to find an optimal efficiently decodable code.  Lastly, the high
 yield of recurrence method when coupled with our adaptive QLDPC code implies
 that our strategy of adaptively combining a high rate QLDPC code with another
 compatible lower rate QLDPC code together with dropping a few qubits in case
 the tentative decoding cannot be found is reasonably effective.

\section{Conclusions And Outlook} \label{Sec:Conc}

 In conclusion, we have pointed out the need to replace universal random
 hashing and breeding methods by an efficiently decodable code in order to make
 an entanglement distillation scheme practical.  As a pilot study, we have
 investigated the yields of the recurrence method using various efficiently
 decodable codes as substitutions for random hashing or breeding methods to
 distill EPR pairs from a collection of Werner states $W_{p_0}$.  We find that
 among the codes we have studied, the best yield over almost all values of
 $p_0$ is achieved by an adaptive QLDPC code with the possibility of dropping a
 few low confidence qubits in the worst case scenario that the tentative
 decoding cannot be found.  Our finding shows that adaptive QLDPC codes are
 useful resources in quantum information processing.

 A number of followup researches along this line have to be done.  For
 instance, both the yields of other practical bipartite and multipartite
 entanglement distillation schemes and the choice of multiple levels of
 adaptive QLDPC codes require thorough investigations.

\section*{Acknowledgment}
 Valuable discussions with C.-H. F. Fung is gratefully acknowledged.
 This work is supported by the RGC grant number HKU701004 of the HKSAR
 government.  We would like to thank the Computer Center of HKU for their
 helpful support in providing the use of the HPCPOWER System for some of the
 simulations reported in this paper.

\bibliographystyle{IEEEtran}
\bibliography{qc43.4}

\begin{thebibliography}{10}
\providecommand{\url}[1]{#1}
\csname url@samestyle\endcsname
\providecommand{\newblock}{\relax}
\providecommand{\bibinfo}[2]{#2}
\providecommand{\BIBentrySTDinterwordspacing}{\spaceskip=0pt\relax}
\providecommand{\BIBentryALTinterwordstretchfactor}{4}
\providecommand{\BIBentryALTinterwordspacing}{\spaceskip=\fontdimen2\font plus
\BIBentryALTinterwordstretchfactor\fontdimen3\font minus
  \fontdimen4\font\relax}
\providecommand{\BIBforeignlanguage}[2]{{%
\expandafter\ifx\csname l@#1\endcsname\relax
\typeout{** WARNING: IEEEtran.bst: No hyphenation pattern has been}%
\typeout{** loaded for the language `#1'. Using the pattern for}%
\typeout{** the default language instead.}%
\else
\language=\csname l@#1\endcsname
\fi
#2}}
\providecommand{\BIBdecl}{\relax}
\BIBdecl

\bibitem{BBP+96a}
C.~H. Bennett, G.~Brassard, S.~Popescu, B.~Schumacher, J.~A. Smolin, and W.~K.
  Wootters, ``Purification of noisy entanglement and faithful teleportation
  over noisy channels,'' \emph{Phys. Rev. Lett.}, vol.~76, pp. 722--725, 1996.

\bibitem{BDSW96a}
C.~H. Bennett, D.~P. DiVincenzo, J.~A. Smolin, and W.~K. Wootters, ``Mixed
  state entanglement and quantum error-correcting codes,'' \emph{Phys. Rev. A},
  vol.~54, pp. 3824--3851, 1996.

\bibitem{Chau02}
H.~F. Chau, ``Practical scheme to share a secret key through a quantum channel
  with a 27.6\% bit error rate,'' \emph{Phys. Rev. A}, vol.~66, pp.
  060\,302:1--4, Dec 2002.

\bibitem{GL03a}
D.~Gottesman and H.-K. Lo, ``Proof of security of quantum key distribution with
  two-way classical communications,'' \emph{IEEE Trans. Inf. Theory}, vol.~49,
  pp. 457--475, 2003.

\bibitem{MS00}
E.~N. Maneva and J.~A. Smolin, ``Improved two-party and multi-party
  purification protocols,'' \emph{AMS Contemporary Mathematics Series}, vol.
  305, pp. 203--212, 2002.

\bibitem{LS07a}
A.~W. Leung and P.~W. Shor, ``Entanglement purification with two-way classical
  communication,'' \emph{Quant. Inf. \& Comput.}, vol.~8, pp. 311--329, 2008.

\bibitem{LS07b}
------, ``Adaptive entanglement purification protocols with two-way classical
  communication,'' 2007, quant-ph/0702156v3.

\bibitem{VV05}
K.~G.~H. Vollbrecht and F.~Verstraete, ``Interpolation of recurrence and
  hashing entanglement distillation protocols,'' \emph{Phys. Rev. A}, vol.~71,
  pp. 062\,325:1--6, 2005.

\bibitem{HDM06}
E.~Hostens, J.~Dehaene, and B.~D. Moor, ``Hashing protocol for distilling
  multipartite \uppercase{C}alderbank-\uppercase{S}hor-\uppercase{S}teane
  states,'' \emph{Phys. Rev. A}, vol.~73, pp. 042\,316:1--13, 2006.

\bibitem{HDM06b}
------, ``Stabilizer state breeding,'' \emph{Phys. Rev. A}, vol.~74, pp.
  062\,318:1--8, 2006.

\bibitem{CEM99}
J.~I. Cirac, A.~K. Ekert, and C.~Macchiavello, ``Optimal purification of single
  qubits,'' \emph{Phys. Rev. Lett.}, vol.~82, pp. 4344--4347, 1999.

\bibitem{BMT78}
E.~R. Berlekamp, R.~J. McEliece, and H.~C.~A. van Tilborg, ``On the inherent
  intractability of certain coding problems,'' \emph{IEEE Trans. Inf. Theory},
  vol.~24, pp. 384--386, 1978.

\bibitem{FCG+88}
J.~Fang, G.~Cohen, P.~Godlewski, and G.~Battail, ``On the inherent
  intractability of soft decision decoding of linear codes,'' \emph{Lect. Notes
  Comp. Sci.}, vol. 311, pp. 141--149, 1988.

\bibitem{BS01}
B.~Sklar, \emph{Digital Communications: Fundamentals and Applications},
  2nd~ed.\hskip 1em plus 0.5em minus 0.4em\relax New York: Prentice Hall, 2001.

\bibitem{MMM04a}
D.~J.~C. MacKay, G.~Mitchison, and P.~L. McFadden, ``Sparse-graph codes for
  quantum error correction,'' \emph{IEEE Trans. Info. Theory}, vol.~50, pp.
  2315--2330, 2004.

\bibitem{DSS98}
D.~P. DiVincenzo, P.~W. Shor, and J.~A. Smolin, ``Quantum-channel capacity of
  very noisy channels,'' \emph{Phys. Rev. A}, vol.~57, pp. 830--839, 1998.

\bibitem{SS07}
G.~Smith and J.~A. Smolin, ``Degenerate quantum codes for \uppercase{P}auli
  channels,'' \emph{Phys. Rev. Lett.}, vol.~98, pp. 030\,501:1--4, 2007.

\bibitem{COT05a}
T.~Camara, H.~Ollivier, and J.-P. Tillich, ``Constructions and performance of
  classes of quantum \uppercase{LDPC} codes,'' quant-ph/0502086.

\bibitem{P88}
J.~Pearl, \emph{Probabilistic Reasoning in Intelligent Systems: Networks of
  Plausible Inference}.\hskip 1em plus 0.5em minus 0.4em\relax San Mateo, CA:
  Morgan Kaufmann, 1988.

\bibitem{MM98}
M.~C. Davey and D.~J.~C. MacKay, ``Low density parity check codes over
  \uppercase{GF}(q),'' \emph{IEEE Comm. Lett.}, vol.~2, pp. 165--167, 1998.

\bibitem{Mac99}
D.~J.~C. MacKay, ``Good error-correcting codes based on very sparse matrices,''
  \emph{IEEE Trans. Info. Theory}, vol.~45, pp. 399--431, 1999.

\bibitem{HC08}
K.~H. Ho and H.~F. Chau, ``An adaptive entanglement distillation scheme using
  quantum low density parity check codes,'' 2008, arXiv:0807.2122.

\bibitem{PC08}
D.~Poulin and Y.~Chung, ``On the iterative decoding of sparse quantum codes,''
  \emph{Quant. Inf. \& Comput.}, vol.~8, pp. 986--1000, 2008.

\bibitem{HI07}
M.~Hagiwara and H.~Imai, ``Quantum quasi-cyclic \uppercase{LDPC} codes,'' in
  \emph{Proceedings of the IEEE International Symposium on Information Theory
  ISIT2007}, IEEE.\hskip 1em plus 0.5em minus 0.4em\relax IEEE, Jun 2007, pp.
  806--810.

\bibitem{HBD08a}
M.-H. Hsieh, T.~A. Brun, and I.~Devetak, ``Entanglement-assisted quantum
  quasi-cyclic low-density parity-check codes,'' \emph{Phys. Rev. A}, vol.~79,
  pp. 032\,340:1--7, 2009.

\bibitem{SAA07a}
S.~A. Aly, ``A class of quantum \uppercase{LDPC} codes constructed from finite
  geometries,'' in \emph{IEEE Global Telecommunications Conference GLOBECOM
  2008}, IEEE.\hskip 1em plus 0.5em minus 0.4em\relax IEEE, 2008, pp.
  1097--1101.

\bibitem{SRK08a}
P.~K. Sarvepalli, A.~Klappenecker, and M.~R\"{o}tteler, ``Asymmetric quantum
  \uppercase{LDPC} codes,'' in \emph{Proceedings of the IEEE International
  Symposium on Information Theory ISIT2008}, IEEE.\hskip 1em plus 0.5em minus
  0.4em\relax IEEE, Jul 2008, pp. 305--309.

\bibitem{AK01}
A.~Ashikhmin and E.~Knill, ``Nonbinary quantum stabilizer codes,'' \emph{IEEE
  Trans. Inf. Theory}, vol.~47, pp. 3065--3072, 2001.

\bibitem{CRSS98a}
A.~R. Calderbank, E.~M. Rains, P.~W. Shor, and N.~J.~A. Sloane, ``Quantum error
  correction via codes over {GF(4)},'' \emph{IEEE Trans. Inf. Theory}, vol.~44,
  pp. 1369--1387, 1998.

\bibitem{LP08}
M.~Leifer and D.~Poulin, ``Quantum graphical models and belief propagation,''
  \emph{Ann. Phys.}, vol. 323, pp. 1899--1946, 2008.

\end{thebibliography}
\end{document}